\documentclass[12pt]{article}
\usepackage{lingmacros}
\usepackage[english]{babel}
\usepackage{graphicx}

\large \baselineskip = 20 true pt

\begin{document}
\begin{center}
{\Large \bf Usage of analytic hierarchy process for steganographic inserts detection in images} \vspace{0.5cm}
\end{center}

\begin{center}
S.V. Belim, D.E. Vilkhovskiy\\
Dostoevsky Omsk State University, Omsk, Russia

 \vspace{0.5cm}
\end{center}

\begin{center}
{\bf Abstract}
\end{center}

This article presents the method of steganography detection, which is formed by replacing
the least significant bit (LSB). Detection is performed by dividing the image into layers
and making an analysis of zero-layer of adjacent bits for every bit. First-layer
and second-layer are analyzed too. Hierarchies analysis method is used for making decision
if current bit is changed. Weighting coefficients as part of the analytic hierarchy process
are formed on the values of bits. Then a matrix of corrupted pixels is generated.
Visualization of matrix with corrupted pixels allows to determine size, location
and presence of the embedded message. Computer experiment was performed. Message was embedded
in a bounded rectangular area of the image. This method demonstrated efficiency even
at low filling container, less than 10\%. Widespread statistical methods are unable to detect
this steganographic insert. The location and size of the embedded message can be determined
with an error which is not exceeding to five pixels.

\section{Introduction}

The simplest and most common method of embedding steganographic inserts is a substitution
of the least significant bits(LSB- substitution) \cite{b1}. The basic idea of the method is
to replace from one to four the least significant bits in bytes of image color representation
pixels. The least visible variant is the replacement of the blue component of the color,
because that is associated with the peculiarities of the human eye color perception.
This method is used both alone and as part of a more complex methods. Despite the simplicity
of the algorithm formation steganographic insert, the problem of detection without additional
information is quite complex. To date, there is no method that can determine with certainty
the existence and the dimensions of the steganographic insert in any container. Most methods
are statistical in nature and are based on the assumption that the change in the statistical
properties of the image bits by placing it in the inserted information. Known to date methods
are effective when steganographic container is filled not less than 50\% \cite{b2}.

In \cite{b3} detection of stenographic insertion is performed on the assumption of change in
correlation between adjacent pixels. The proposed by the authors  method is that nearest
neighbors of each pixel are considered. From the analysis of the surrounding pixels we can make
predictions about the value of the center pixel and compare with current value.

In \cite{b4} an algorithm for detecting steganographic inserts with using templates
for surrounding pixels. Template building is also based on the assumption of a strong
correlation between pixels of the original image. Correlations between pixels is also used
in \cite{b5} to build the statistical method for detecting steganographic inserts. A similar
statistical method based on the the value of correlation between adjacent pixels is proposed
in \cite{b6}. In \cite{b15} there is a generalized method for determining the length of the steganographic inserts by combining multiple detectors.

Using the autoregressive model for the detection of hidden messages, as well as an assessment
of their relative lengths is proposed in \cite{b7,b8}.

Thus, the main steganalysis objectives today are the fundamental discovery of the presence of
a hidden insert and, if possible, the length determination of it. The purpose of this article
is a developing an algorithm for deciding whether a particular bit is spoofed. That is not
simply to determine the presence of steganographic insert, and, if possible, its definition.

\section{Formulation of the problem}

We will analyze the images, which can have embedded information in the form of steganography
inserts in the least significant bit of the blue component. In this case we start from the
two assumptions. First, we assume that we don't know if there is any steganographic. Secondly,
it is not known in advance about number of embedded bits and their geometric position in the
image. The problem is posed to detect steganographic insert and determining the maximum number
of pixels, which have spoofed the LSB of the blue component. The second assumption significantly
complicates the task, because there can be a situation when pixels are replaced in all zero-layer
bits of blue components. In this case, analysis  of the zero layer of image does not give any
information. At the same time it is not known in advance whether to allow the zero-layer analysis
to make any conclusions. In this case the analysis requires higher layers. We will rely on the
assumption that basic regularities of the image gradually changes from one layer to another.
Therefore detected regularity in one layer must be repeated with high probability in the
surrounding layers.

We will find pixels with substitution in zero bits with separately analyzing zero layer and next
three layers. In the future, we will build a chart of the results of these two algorithms, and
adopting a general solution.

Let the $k$-th layer of the blue component of the original image is defined as a binary color
matrix $B_{ij}^{(k)}$ , and the coordinates of the embedded information are set in the form
of a matrix $R_{ij}$. In this case, $R_{ij}=1$, if there is a substitution LSB blue components
of the corresponding pixel and $R_{ij}=0$, if there is no substitution. As a result of embedding
steganographic insert, instead of the zero layer $B_{ij}^{(0)}$ matrix will be formed
$A_{ij}^{(0)}$. The problem is reduced to the most accurate restoration of matrix $R_{ij}=0$
from analysis of the matrices $A_{ij}^{(0)}$, $B_{ij}^{(1)}$, $B_{ij}^{(2)}$, $B_{ij}^{(3)}$.

\section{Application of the analytic hierarchy process to identify spoofed bits}

Let's apply the analytic hierarchy process \cite{b9} for a decision on the substitution
of the bit. This requires to formulate alternative solutions, from which selection is performed
and also criteria for analyzing alternatives. As mentioned in the statement of the problem it's
necessary to identify the pixels with LSB substitution. Therefore only one of the possible two
solutions denoted hereafter or $Y$, if there is a substitution in LSB of given pixel, or $N$,
if pixel was not changed.

First, we construct a system to identify the substitution of bits based on the analysis of the zero layer. For this we will perform sequential pass over all bits of zero layer and make an analysis of the nearest neighbors of each of them. We distinguish three criteria:

$K_1$ -- adjacent bits on the sides have the same value as the analyzed or different from it.

$K_2$ -- the corners adjacent bits have the same value as the analyzed or different from it.

$K_3$ -- bit deviation from the average value of surrounding eight bits.

The first two criteria allow to detect extended regions of the same color on image. The third criterion is needed to identify areas with a gradient. Thus, we obtain a two-level hierarchical tree of alternatives which is shown in Figure 1. The final decision is indicated by $R$.

\begin{figure}[ht]
\centering
\includegraphics[width=0.6\textwidth]{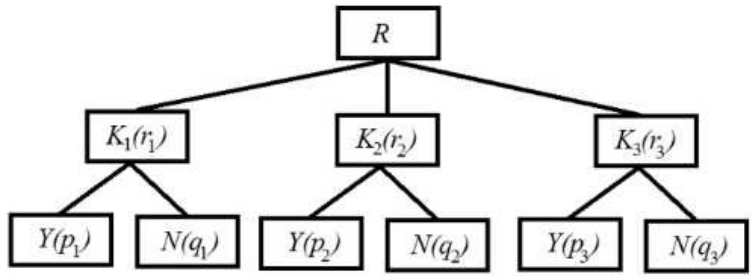}
\caption{Hierarchy of criteria to determine the substitution bit from zero layer analysis.}
\label{fig1}
\end{figure}

For the application of the analytic hierarchy process it is required to determine the relative
weights of the criteria $r_i$ ($i=1,2,3$), and weight solutions within one criteria and weight
solutions within one criteria $p_i$ and $q_i$ ($i=1,2,3$). We will assume that the criterion
$K_1$ is more important than $K_2$ in $n$ times, and criterion $K_2$ is more important than
$K_3$ in $k$ times. Also, we assume the presence of transitivity, this means that $K_1$ is more
important than $K_3$ in $nk$ times. Then coherent matrix of pairwise comparisons will look like:
\begin{center}
\begin{tabular}{|c|c|c|c|}
  \hline
   & $K_1$ & $K_2$ & $K_3$ \\
  \hline
  $K_1$ & 1 & $n$ & $kn$ \\
  \hline
  $K_2$ & $1/n$ & 1 & $k$ \\
  \hline
  $K_3$ & $1/(kn)$ & $1/k$ & 1 \\
  \hline
\end{tabular}
\end{center}

From this matrix the weighting coefficients can be obtained with standard methods \cite{b9}:
\[
r_1=\frac{nk}{nk+k+1},\ \ r_2=\frac{k}{nk+k+1},\ \ r_3=\frac{1}{nk+k+1}.
\]

With classical usage of the analytic hierarchy process pairwise comparisons are determined
on the basis of expert evaluations. In our approach, we use some objective indicators instead
of expert evaluations, which are determined by the number. In particular, constraints
on the values $n$ and $k$ we will determine from consideration trivial examples further.
The most suitable values of these parameters will be obtained from computer experiment.

Let's turn to the definition of the weighting factors in each of the criteria. We begin with
$K_1$. Let four bits of contacting with $x$ have the same value, then the solution $N$ has more
weight as compared with $Y$ (namely analyzed bit is not spoofed) in $x/(4-x)$ times.
Writing the matrix of pairwise comparisons and making the necessary changes, we are getting the
values of the coefficients $p_1=(4-x)/4$, $q_1=x/4$. Similarly for the criterion $K_2$. Let four
bits in contact with current bit only at the vertices have the same value. Then, the weight coefficients will be changed to $p_2=(4-y)/4$, $q_2=y/4$.

For the calculation of weight coefficients according to the criterion let't assume that value of
analyzed bit and the average value of the surrounding bits is $c_0$. The following arguments
are used to find weight coefficients. Let the solution $N$ has more weight than $Y$ in $a$ times,
where the value $a$ depends on the absolute value of the deviation value of the bit, $c$ depends
on the average values of surrounding bits $c_0$ ($dc=|c-c_0|$). Then, the weight coefficients
will have the form:

\[
p_3=\frac{1}{a+1},\ \ q_3=\frac{a}{a+1}.
\]

Let's consider extreme cases. If the current bit is equal to the average value of the surrounding
bits ($dc=0$) we will assume that this bit is not spoofed, the coefficients of this bit will have
this values $p_3=0$, $q_3=1$. If bit has maximum difference from the surrounding ($dc=1$), then
we will think that it was spoofed, i.e $p_3=1$, $q_3=0$. Therefore, when $dc=0$ there should be
$a\rightarrow \infty$. So, if $dc=1$ then $a=0$. These conditions are satisfied the following
expression:
\[
a=\frac{1}{dc}-1.
\]
Weighting coefficients will have the following values:
\[
p_3=dc,\ \ q_3=1-dc.
\]
For a final decision it is necessary to calculate values:
\[
P(Y)=r_1p_1+r_2p_2+r_3p_3,\ \ P(N)=r_1q_1+r_2q_2+r_3q_3.
\]

If $P(Y)>P(N)$, then we will make a decision that $R=Y$, this means that bit is spoofed,
otherwise, if $P(Y)\leq P(N)$, then we will make a decision that $R=N$, this means that bit
is not spoofed.

We will extend the proposed method of bit analysis with a comparison of three overlying layers.
We will consider bits in each layer, which lies over the data and eight nearest adjacent bits.
In the future, this set of bits will be called as window in the corresponding layer.
Let's introduce the criteria for making  decision based on analysis of $k$-th layer ($k=1,2,3$):

$K_1^{(k)}$ -- adjacent bits on both sides in the window of $k$-th layer have the same value as
the analyzed bits of zero-layer or different from it.

$K_2^{(k)}$ -- the corners bits on both sides in the window of $k$-th layer have the same value
as the analyzed bits of zero-layer or different from it.

$K_3^{(k)}$ -- deviation of bit in zero-layer from the average value of bits in window of $k$-th
layer. Three-level hierarchical tree of alternatives is shown in figure 2. The final decision is denoted $R_1$.

\begin{figure}[ht]
\centering
\includegraphics[width=\textwidth]{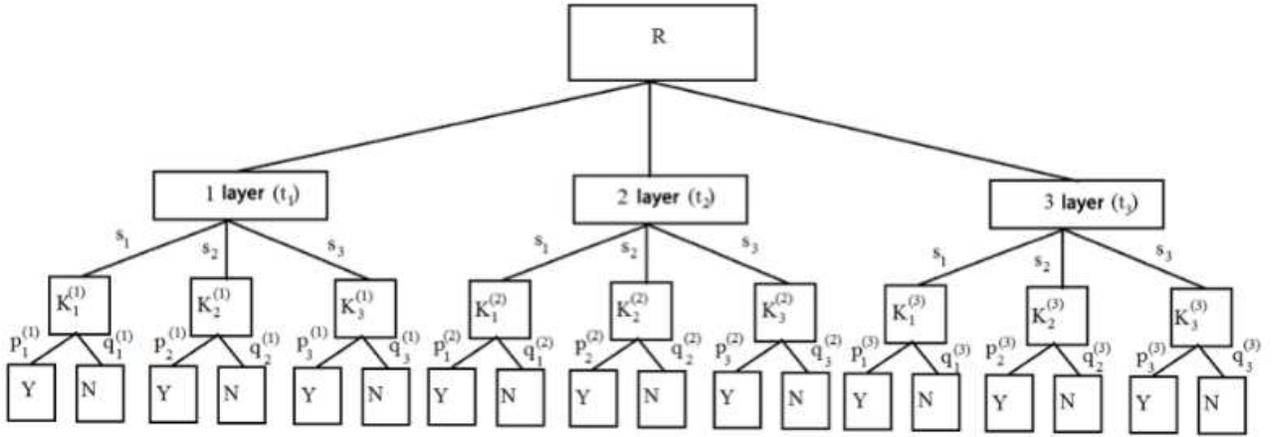}
\caption{Hierarchy of criteria to determine bit substitution from analysis of overlying layers.}
\label{fig2}
\end{figure}

We will assume that the results of the analysis of the first layer is more important than the results of the second twice, and the second layer is more important than the third twice too. Hence, we obtain the values of the weighting coefficients:
\[
t_1=\frac{4}{7},\ \ t_2=\frac{2}{7},\ \ t_3=\frac{1}{7}.
\]

Within the framework of a single layer it is difficult to single out some of the criteria,
so we assume that they are all equal:
\[
s_1=s_2=s_3=\frac{1}{3}.
\]
For the weighting coefficients for two decisions under one criterion we apply an approach which
is similar to used in the zero layer analysis. For the first criterion:
\[
p_1^{(k)}=\frac{4-x^{(k)}}{4},\ \ q_1^{(k)}=\frac{x^{(k)}}{4},
\]
where $x^{(k)}$ is a number of neighbors on each side, with the same value in the window of $k$-layer. For the second criterion:
\[
p_2^{(k)}=\frac{4-y^{(k)}}{4},\ \ q_2^{(k)}=\frac{y^{(k)}}{4},
\]
where $y^{(k)}$ is a number of neighbors on the diagonal, with the same value in the window of $k$-layer. The weighting coefficients of the third criterion:
\[
p_3^{(k)}=dc^{(k)},\ \ q_3^{(k)}=1-dc^{(k)},
\]
where $dc^{(k)}$ -- difference between the bit value from the average value of bits in the window
of $k$-layer.

For making decision it is necessary to calculate values:
\[
P_1(Y)=t_1\left(s_1p_1^{(1)}+s_2p_2^{(1)}+s_3p_3^{(1)}\right)+
t_2\left(s_1p_1^{(2)}+s_2p_2^{(2)}+s_3p_3^{(2)}\right)+
t_3\left(s_1p_1^{(3)}+s_2p_2^{(3)}+s_3p_3^{(3)}\right),
\]
\[
P_1(N)=t_1\left(s_1q_1^{(1)}+s_2q_2^{(1)}+s_3q_3^{(1)}\right)+
t_2\left(s_1q_1^{(2)}+s_2q_2^{(2)}+s_3q_3^{(2)}\right)+
t_3\left(s_1q_1^{(3)}+s_2q_2^{(3)}+s_3q_3^{(3)}\right).
\]

If $P_1(Y)>P_1(N)$, then we make decision $R_1=Y$, i.e. bit is spoofed, otherwise,
if $P_1(Y)\leq P_1(N)$, then we make  decision $R_1=N$, i.e. bit is not spoofed.

\section{The algorithm to detect spoofed pixels}

Let us write the algorithm formally that implements the proposed method. We perform a consistent passage for all pixels in the image. For each pixel we will perform a series of steps:

Step 1. Select size of the window $3\times 3$ in the zero, first, second and third layers.

Step 2. Calculate values $P(Y)$, $P(N)$, $P_1(Y)$, $P_2(N)$.

Step 3. If at least one of the two equations is true $R=Y$ or $R_1=Y$, that we assume that
the bit is spoofed. Then we will add this element to matrix $R_{ij}$ with value 1, otherwise
value is 0.

The output matrix $R_{ij}$ will have list of spoofed pixels. Since the algorithm is performed in a single pass for all pixels and each pixel is performed for a fixed number of steps, then complexity of the algorithm is linear. Also of note is the localization of the data which is needed for making  decision, in a small area around the current pixel, it's easy to make a simple parallelization algorithm with partitioning the image into regions.

\section{Computer experiment and results}

Computer experiment was performed to research the effectiveness of the proposed method to detect
embedded information. The experiments were performed on three types of images: gradient fill, artificial
image of geometric shapes and widely used image "Lena". All images had dimensions $640\times480$
pixels, the color depth is 256 colors. Embedded text in Russian in the form of bit-sequences
in a rectangular area located randomly in the center of the image. There were $9\%$ of changes
in zero-layer of the original image. Figure 3 shows the results of the experiment with
the gradient fill.

\begin{figure}[ht]
\centering
\includegraphics[width=0.9\textwidth]{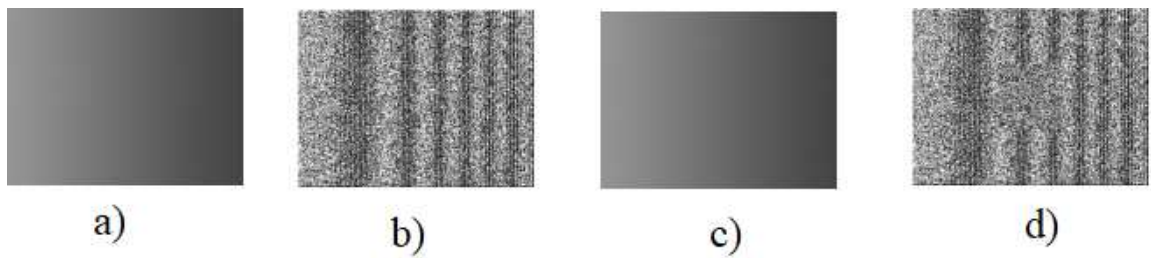}
\caption{Results of the algorithm to detect spoofed pixels on the gradient fill:
a) original image; b) matrix $R_{ij}$ for the original image; c) image with integrated text;
d) matrix $R_{ij}$ for the image with steganography insert.}
\label{fig3}
\end{figure}

As can be clearly seen from a comparison of figures 3a and 3b we can see visible rectangle,
which has embedding. Similar results for the artificial image with a geometric figures are shown
in Figure 4 for the photographic image results are shown in Figure 5.

\begin{figure}[ht]
\centering
\includegraphics[width=0.6\textwidth]{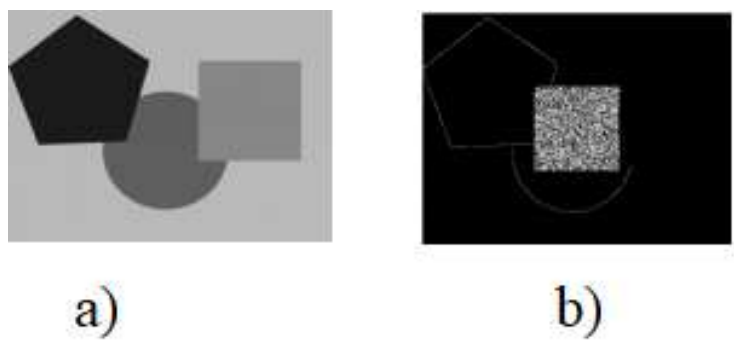}
\caption{Results of the algorithm to detect spoofed pixels on the artificial image with a geometric figures:
a) image with integrated text; b) matrix $R_{ij}$ for the image with steganography insert.}
\label{fig4}
\end{figure}

\begin{figure}[ht]
\centering
\includegraphics[width=0.6\textwidth]{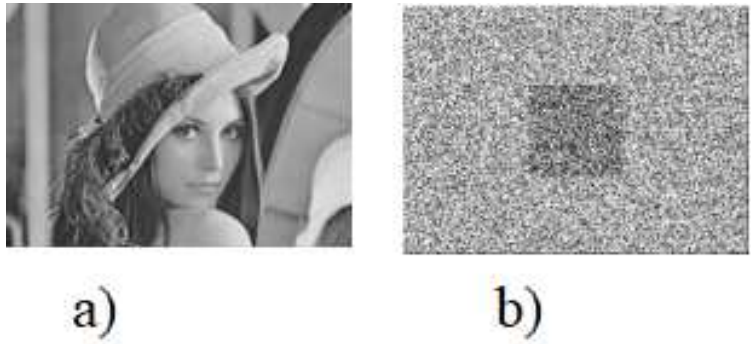}
\caption{Results of the algorithm to detect steganographic insert on the photographic image:
a) image with integrated text; b) matrix $R_{ij}$ for the image with steganography insert.}
\label{fig5}
\end{figure}

Both figures have clearly visible area, which was built by inserting hidden message.

\section{Conclusion}

Thus, the proposed algorithm in this article allows us not only to detect the presence
of steganography insertion into the image, but also it allows to determine with sufficient
accuracy location and volume of steganography insertion. Unlike presently common algorithms,
in this proposed method there was not any statistical approach. It should be noted that
the all used images images with steganographic insertions easily pass the Chi-square test
which detects embedded messages in them. This method requires further research for
the development of algorithms for the analysis of the resulting matrix $R_{ij}$.



\begin{thebibliography}{99}

\bibitem{b1}
Adelson E. Digital Signal Encoding and Decoding Apparatus. U.S. Patent. 1990, N. 4,939,515.
\bibitem{b2}
Westfeld A., Pfitzmann A. Attacks on Steganographic Systems. Breaking the Steganographic
Utilities EzStego, Jsteg, Steganos and S-Tools and Some Lessons Learned. Lecture Notes
in Computer Science. 2000, 1768, pp. 61-–75.
\bibitem{b3}
Zhang J., Xiong F. Steganalysis for LSB Matching Based on the Dependences Between Neighboring
Pixels. Journal of multimedia. 2012, V.7, 5, pp. 380--385.
\bibitem{b4}
Lerch-Hostalot D., Megias D. LSB Matching Steganalysis Based on Patterns of Pixel Differences
and Random Embedding. Computers and Security. 2013, V. 32, pp. 192--206.
\bibitem{b5}
Xia Zh., Wang X., Sun X., Wang B.  Steganalysis of least significant bit matching using
multi-order differences. Security and Communication Networks. 2014, V. 7, 8, pp. 1283–-1291.
\bibitem{b6}
Guan Q., Dong J., Tan T. An effective image steganalysis method based on neighborhood information
of pixels. 18th IEEE International Conference on Image Processing. Brussels,Belgium 2011,
pp. 2777-–2780.
\bibitem{b7}
Andrew D., Ker A. General Framework for Structural Steganalysis of LSB Replacement. M. Barni et al. (Eds.): IH 2005, LNCS 3727, pp. 296-–311.
\bibitem{b8}
Bhattacharyya S., Sanya G. Steganalysis of LSB Image Steganography using Multiple Regression and Auto Regressive (AR) Model. Int. J. Comp. Tech. Appl. 2011, V. 2(4), pp. 1069-–1077.
\bibitem{b9}
Saaty T.L. Relative Measurement and its Generalization in Decision Making: Why Pairwise
Comparisons are Central in Mathematics for the Measurement of Intangible Factors The Analytic
Hierarchy. Network Process. Review of the Royal Spanish Academy of Sciences, Series A,
Mathematics, 2008, V. 102 (2), pp. 251-–312.




\end{thebibliography}
\end{document}